# BlockSim-Net: A Network Based Blockchain Simulator

*Nandini Agrawal[1], Prashanthi R[1], Osman Biçer[2], Alptekin Küpçü[2]*

1. Ashoka University (Work done while at Koç University)
{nandini.agrawal,prashanthi.r}@alumni.ashoka.edu.in

2. Koç University
{obicer17,akupcu}@ku.edu.tr

**Abstract**

*Since its proposal by Eyal and Sirer (CACM '13), selfish mining attack on proof-of-work blockchains has been studied extensively in terms of both improving its impact and defending against it. Before any defense is deployed in a real world blockchain system, it needs to be tested for security and dependability. However, real blockchain systems are too complex to conduct any test on or benchmark the developed protocols. Some simulation environments have been proposed recently, such as BlockSim (Maher et al., '20). However, BlockSim is developed for the simulation of an entire network on a single CPU. Therefore, it is insufficient to capture the essence of a real blockchain network, as it is not distributed and the complications such as propagation delays that occur in reality cannot be simulated realistically enough. In this work, we propose BlockSim-Net, a simple, efficient, high performance, network-based blockchain simulator, to better reflect reality.*

## 1. Introduction

Proposed by Nakamoto [1], blockchain has applications, in cryptocurrencies, certificate transparency, supply chain management, governmental services, etc., thanks to its transparency and immutability features. They are maintained by peer-to-peer (P2P) networks, composed of nodes called "miners". Miners are incentivized by a competition through a process of mining, i.e., a trial-and-error process conducted via invested computational resources.

Nakamoto's security assumption for Bitcoin was that as long as a majority of the computational resources belongs to honest miners, the system will award each miner their fair share. In other words, the reward that each miner receives will be proportional to their invested resources or "hashing power". However, in 2018, Eyal and Sirer [2] showed that an adversarial miner can obtain more than their fair share even without having a majority of the total hashing power. Since then, there has been extensive research to optimize this attack [3-6], and to defend against it [7,8]. The proposal of these improvements clarified the need for a dependable environment, to conduct the tests. However, real blockchain systems are too large and complex for these purposes. In recent times, some general simulation environments have been developed [9-11]. We focus on BlockSim [11], which is proposed for simulation on a single CPU. While BlockSim might suffice for certain security tests, a network based simulation would be more decentralised and comparable to a real blockchain. Such a simulator can also help simulation of other attacks and issues associated with the network itself, e.g., network randomness due to propagation delays and oracle and bold mining attacks [12]. Therefore, in this work, we propose *BlockSim-Net*, a simple, easy-to-use, highly-efficient, network based blockchain simulator for use in blockchain security research. Additionally, our simulator can also be used for beta-testing and optimisation benchmarking.

### 1.1. Our Contributions

We build on top of an existing work, BlockSim [11] which simulates a blockchain on a local system. The contribution of this work is two-fold:

1. *Local to Network:* Our main contribution is converting the BlockSim's [11] local simulation to a network simulation. As opposed to a local simulation, a network simulation is more distributed, decentralised, and comparable to a real blockchain, since multiple miners from different locations can be involved in simulation of a blockchain network together. Such a simulator can also be used to effectively simulate network-based attacks.
2. *Efficiency:* Since simulation on the network comes with associated heavy communication costs, our work also includes multiple optimisations. The main optimisations have been mentioned below:
   A. *Switching to another miner's blockchain when a block is received*: Since [11] simulates on a single system, miner nodes can easily access the local blockchains of other miner nodes.

      Over a network, however, to switch to another miner's blockchain, a node would need to receive the entire blockchain of that miner. This would mean high communication costs. Thus, instead of that, we propose the use of a local dictionary that stores all the blocks that the miner receives during the simulation. Most of the time, using this local dictionary, miners will be able to recreate the chain that was supposed to be received with minimal communication.

   B. *Consensus*: In [11], since the entire simulation happens on a single system, the main thread can access the longest blockchains of all miners at the end of the simulation to decide on the longest blockchain. However, over a network of miners, each miner would have to send their longest local blockchain to the admin server for consensus (refer 2.3.3). However, this propagation of entire blockchains would be very costly in terms of communication complexity. Thus, instead of doing that, in our work, we propose that miners only send the last block of their longest local blockchain to the admin server for consensus. We elaborate on why the last block is enough for consensus in Section 2.3.3.

### 1.2. BlockSim-Net vs. Other Simulations

There has been a lot of work done related to blockchain simulators. Some help you analyse certain characteristics and properties of a bitcoin like network [9-10], whereas some are more general and bootstrap an entire network in a test environment where the parameters are initialised by the user like [11]. BlockSim-Net can be used to simulate both Bitcoin or Ethereum test network or to simulate certain features, attacks, and defences. It is built on top of BlockSim, but instead of being a single PC simulator, it is over a network. While [11] is highly scalable, BlockSim-Net is better for security analyses because it is a closer approximation of an actual blockchain network. A qualitative comparison between Blocksim-Net and [11] has been shown in Table 1.

### 1.3. Related Work

#### 1.3.1. Selfish Mining (SM) Attacks

In the original bitcoin paper, Satoshi Nakamoto [1] works on the assumption that as long as the majority of the hashpower belongs to honest miners, every miner gets a reward that is proportional to their work. However, Eyal and Sirer [2] have introduced the concept of selfish mining which allows a miner more than her fair share, even when the honest majority assumption holds true. The primary idea behind selfish mining is that a seemingly-honest miner can deviate from the honest protocol by populating a private chain until it becomes longer than the existing longest chain in the network. By extending on a private chain, a miner can potentially eliminate blocks in the network and get rewarded more than one's fair share.

#### 1.3.2. Selfish Mining (SM) Defences

A couple of selfish mining defenses have also been proposed by various works [2,7,8]. In the original bitcoin implementation, in case multiple chains with the same number of blocks occur, miners choose the one that they receive first. However, [2] proposes choosing one of the chains uniformly at random. This defense prevents attackers with less than 25% of the total hashpower of the network from obtaining high rewards. However, the later attack proposal by [3] on Optimal Selfish Mining reduces this value to 23.2%. [8] proposes the publish or perish scheme and this takes the value back to 25%.

Given this trend of proposals of attacks and defenses, blockchain security research needs to constantly work to improve with better defenses. In this scenario, there arises a need for simulators that can simulate a real network as realistically as possible in order for defenses to be tried and tested. Here is where we think our simulator BlockSim-Net would be beneficial. BlockSim-Net can be used to simulate proposed attacks and defenses.

## 2. Implementation

BlockSim-Net is intended to be a simulation of a blockchain on a real network with different servers acting as miners. For implementation, we have split the codebase into two sections:

|  | Blocksim | Blocksim-Net |
|---|---|---|
| Distributed | ✗ | ✓ |
| Propagation of Blocks | Simulates propagation of blocks | Actual propagation of blocks |
| Structure | Simulates all miners in a single system | Can be simulated through multiple servers **over a network** |
| Scalability | Depends on the computation power of the single CPU | Depends on the availability of computational resources |

*Table 1:* A qualitative comparison between BlockSim [11] and BlockSim-Net.

1) The admin server; and 2) The miner. The next section discusses these in detail.

### 2.1. Structure

Structurally, our blockchain simulator has two aspects—the miner and the admin server. All miners on the network can deploy the miner code at their ends and run a simulation collectively. The admin server—discussed extensively in the following section—runs on one node during the simulation.

#### 2.1.1. The Admin Server

For our purposes, we have used an admin server, **AS**, that acts as a common point of contact to each of the miners. Every node connects to this server in the beginning of a simulation. **AS**, having received connections from the miners, collects information about them – miner ID, hashpower, IP, the port on which each of the miners will host their server socket during the simulation. We will call this information *miner-info*. Being the common point of contact for each of the miners, **AS** also shares the information of other miners in the network. Each node can communicate with each other for the purposes of the simulation. Although, in reality, such an admin server is not present in a blockchain, here we need it for the sake of efficiency, and utilize it only in the start and end of the simulation.

We emphasize that **AS only initiates** the miners for the simulation and has no role in the creation and exchange of blocks in the simulation. **AS** assigns an ID to each of the miner nodes in order to make identification of the miner nodes easier. Once each of the nodes has been provided with *miner-info*, they are ready to run the simulation independently without any help from **AS**. *miner-info* includes miner ID, total hashpower of all the nodes in the network, the IP and the port on which each of the miners will host their server socket during the simulation.

Apart from communicating information about other miners in the network with each node, **AS** is responsible for the creation of the genesis block. It is also responsible for creating a pool of transactions that are to be used by the miners for the duration of the simulation. As all miners have access to the genesis block and a transaction pool (with common transactions) in an actual network, the **AS** provides the genesis block and a pool of transactions to all the miners at the beginning of a simulation as a common starting point.

The next time that **AS** figures in the simulation is at the time of **consensus,** i.e., once the simulation is complete. While in a real blockchain, consensus happens as a result of communication between the miners themselves, here, it is important to remember that BlockSim-Net performs a *timed- simulation*; a simulation has a start-time and an end-time. Due to this property of a simulator, there arises a need for an entity that can make up for a potential abrupt halt in the network. Thus, when the simulation end-time is reached, we use an entity like **AS** to ensure that the local blockchains of all miners are complete and uniform.

*2.1.2.    The Miner*

The role of a miner is to create blocks on the network and propagate those to the other miners on the network. A miner also listens to the other miners on the network for blocks. Upon receiving blocks from other miners, a miner updates/maintains their blocks, based on some predefined rules. These rules differ from one another based on the type of blockchain or coin.

Initially, each one of the miners connect to the admin server, **AS**. They learn *miner-info* from **AS**. As mentioned in the previous section, *miner-info* includes – miner ID, total hashpower of all the nodes in the network, the IP and the port on which each of the miners will host their server socket during the simulation. Once this basic information is acquired, the simulation timer starts running.

As soon as the simulation begins, **AS** creates a genesis block and a transaction pool and sends them to the miner nodes. Each miner node upon receiving the genesis block and a pool of transactions can start creating a new block at depth 1 (considering the genesis block is at depth 0). Each miner also parallely listens for blocks from other miners on the network.

**2.2.    Initialization**

One of the main responsibilities of **AS** is to provide basic information to the miners in the network. This information includes the miner ID to uniquely identify a miner, IP addresses and ports of all the miners on the network so that they can communicate with each other, and the total hashpower of the network which is computed by the admin server by adding the individual hashpowers of each miner. This basic information is referred to as *miner-info* in our paper. Additionally, at the start of the simulation, **AS** also provides a pool of transactions to be used for a simulation period and the genesis block to the miners at the start of each simulation.

On the miner end, each of the miners randomly initialise themselves with a hashpower (a random sample from 0-30). After receiving their IDs from the admin server, the miners also initialise themselves with that value.

**2.3.    Simulation**

At the start of the simulation, each miner gets a genesis block and a pool of transactions from **AS**. After this, they start mining on top of the genesis block. The created blocks are stored in a 'create queue'. Parallely, the miners also listen for new blocks from other miners on the network. The received blocks are stored in a 'receive queue'. Refer to *Figure 1* for the summary of the entire simulation.

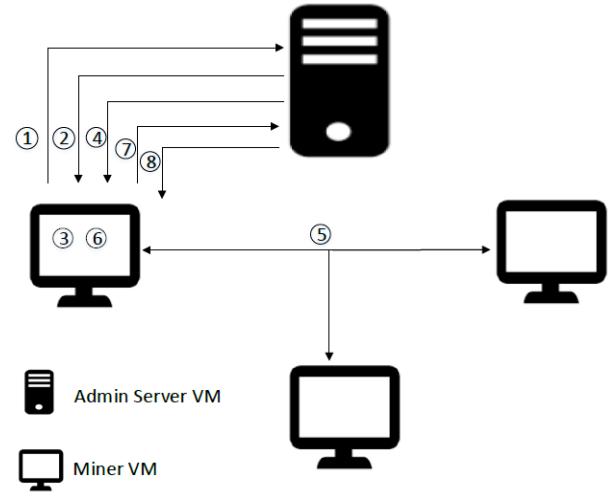

*Figure 1*: BlockSim-Net Simulation Flow: **1.** Miner connects to the admin server. **2.** Miner gets basic information from the admin server. **3.** Miner starts simulation. **4.** Miner gets the Genesis Block and Transaction Pool from the admin server at the start of the simulation. **5.** Miner starts the process of creating blocks and listening to blocks created by other miners on the network parallely. With every new block (either created or received), miner may update their local blockchain. **6.** At the end of simulation, Miner replaces the empty blocks (if any) in its longest chain with the blocks received. **7.** Miner sends the last block of their longest chain to the admin server for consensus. **8.** Based on the results of consensus, the miner gets the longest chain from the admin server (if their chain isn't the longest).

In order to simulate proof of work in our blockchain network, we use the concept of **"events"** similar to [11]. In our implementation, each miner uses their (randomly generated) hash power, *h*, in conjunction with the value of total hashpower obtained from **AS** to calculate the time (in seconds) a miner would take to create a block. Thus, **creation of a block**, in our context, means creation of a block object that is scheduled to be sent out into the network only when its blocktime (as calculated) is reached (or exceeded). As mentioned above, once created, these block objects are added to 'create queue'.

The miner, alternatively checks for blocks in the 'create_queue' and the 'receive_queue'. When a block is created, the miner appends it to its own blockchain if the certain conditions are met. On the other hand, if new blocks have been received, the miner checks if the blocks impose any further action to be taken on our blockchain.

*2.3.1.    Updating local blockchain: Rules and action to be taken*

Any update to the local blockchain of a miner happens based on some predefined rules. A miner node may have to update its local blockchain in two cases: a. when it creates a block, b. when it receives a block from another miner node in the network. These rules will be discussed in detail in this section.

a. When it creates a block:

As discussed above, in this paper, 'creation of a block' actually corresponds to scheduling of an **event**. When a block object is created, it is appended to the miner's 'create queue'. Any action on the created block is only taken once the block object's blocktime has been reached or exceeded. As we have also discussed above, a miner needs to constantly check the blocks in their 'create queue' to see if the blocktime of the block with the earliest timestamp in the queue, *next_create_block*, has been reached (or exceeded). Once the blocktime of *next_create_block* has been reached (or exceeded), the miner looks at the last block, *last_block* in their local blockchain. If the depth of *last_block* is less than that of *next_create_block*, the miner appends *next_create_block* to their local blockchain. After this process, *next_create_block* is removed from the queue. The miner can now start mining on top of *next_create_block*, which is now the last block of the updated blockchain. If a block with the same depth as *next_create_block* is received, it will be added to the miner's uncle chain.

b. When it receives a block:

When a miner receives a block from another miner in the network, the block is first added to the 'receive queue'. During the simulation, the miner fetches the block with the earliest timestamp, *next_receive_block*, from the 'receive queue'. If the depth of this block is greater than that of the last block in the miner's local blockchain, the miner needs to update their blockchain. If not, *next_receive_block* is added to the uncle chain.

If there is a need to update one's local blockchain, the following rules have to be followed:

- If a received block is built on top of the last block of the miner's longest local blockchain, the miner appends it to their longest chain. If they were already trying to mine a block at the same depth as the received block, they will discard their block. The miner, having updated their local blockchain, can now start mining on top of the received block.

- However, if the received block is not built on top of the last block of the miner's longest chain, the miner looks at its depth. If the depth of the received block is less than or equal to the last block of the miner's longest local blockchain, the miner adds the block to its uncle chain. However, if the depth of the received block is greater than that of the last block of the miner's longest local blockchain, the miner needs to switch to the chain of the miner from whom the block was received.

*2.3.2. Switching to another miner's blockchain*

A simulation entails multiple exchanges of blocks. However, not all of the blocks involved -- created or received, make it to the longest chain of a miner immediately. However, because of consensus, many of these blocks make it to the miner's longest chain at some point with a considerable probability. Thus, if some blocks are discarded because they do not make it to the miner's longest blockchain now and they make it to the miner's longest blockchain at a point in the future, it does not make sense to discard them.

In our implementation, the miners locally store all the blocks that they create and receive from other miners on the network in a dictionary with the block IDs as keys and block objects as values. Thus, when a miner has to switch to another miner's chain, they trace back from the latest received block up to the genesis block. If the miner does not have some blocks in their local dictionary due to potential delays, we put temporary empty blocks up to the genesis block. We then try to replace the empty blocks with the correct blocks, if available in the dictionary, once the simulation ends.

*2.3.3. Consensus*

At the end of the simulation, each miner sends the last block of their longest local blockchain to the admin server. The admin server determines the miner with the longest chain from the depth of these blocks. In case more than one miner has the longest blockchain, it considers the miner whose last block was created the earliest.

Once the admin server determines the miner with the longest chain, the admin server communicates with this miner to obtain their longest local chain. After that, the admin server sends it to all the other miners. If the miner with the longest chain has empty block objects as part of the chain which could not be replaced (due to the unavailability of those blocks), that simulation is discarded.

## 3. Results

For a realistic simulation, we have deducted the mining statistics of the top 6 pools (F2Pool, Poolin, BTC.com, AntPool, Huobi.pool, and ViaBTC) for the time interval between (and including) 22 Oct 2019 and 21 Oct 2020 from btc.com [13]. We feed this information as the hashing powers for 7 miner nodes (6 pools and others) in our 2 simulations S1 and S2 whose results are presented in Table 2. Our simulation code is available at [14]. The specific information of how these simulations have been set is given below:

(i) Simulation S1:
Average block interval = 12.42 seconds
Simulation time = 1000 seconds
Total number of blocks mined: 75

(ii) Simulation S2:
Average block interval = 12.42 seconds
Simulation time: 1500 seconds
Total number of blocks mined: 117

From Table 2, it can be seen that both our simulation results show the proof of work guarantee: every miner gets a fair share (total block percentage) proportional to their hashing power. Also, Figure 2 and Figure 3 provide a pie chart of the real hash rates and S2 for comparison.

| Mining Pools and hash percentage | S1 | S2 |
|---|---|---|
| | % of total blocks mined | % of total blocks mined |
| F2Pool (17.0%) | 18.67 | 19.66 |
| Poolin (15.8%) | 16.00 | 17.09 |
| BTC.com (12.9%) | 10.67 | 10.26 |
| AntPool (11.0%) | 10.60 | 9.40 |
| Huobi.pool (6.6%) | 8.04 | 7.69 |
| ViaBTC (6.3%) | 8.00 | 1.71 |
| Others (30.4%) | 28.00 | 34.1 |

*Table 2: Block percentage for major miners in two types of Blocksim-Net simulations.*

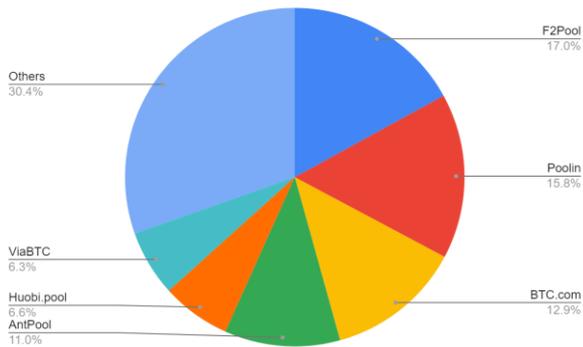

Figure 2: Real Bitcoin pool distribution of hash rates according to [13].

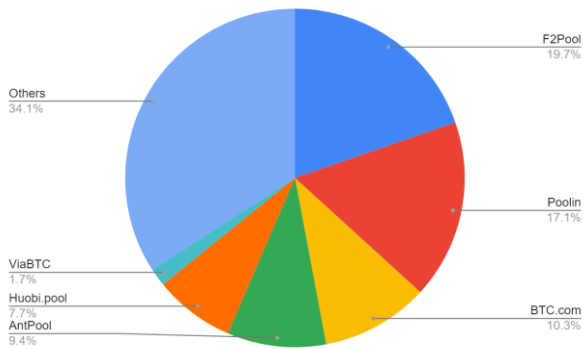

Figure 3: Blocksim-Net simulation S2: Distribution of percentage of blocks mined.

### 4. Conclusion

In this paper, we propose BlockSim-Net, a simulation framework for blockchain technologies like Bitcoin and Ethereum, on a real network with multiple communication optimisations. Our codebase in Python, i.e., an extension of the implementation of Blocksim, has been divided into two aspects - the admin server and the miner.

Future work is needed to improve on the scalability of the system. Furthermore, while BlockSim-Net provides a codebase for Bitcoin and Ethereum networks, our work could potentially extend to other blockchain systems and simulate possible attacks and defence mechanisms on these systems.